\documentclass[a4paper]{jpconf} 

\usepackage{dcolumn}
\usepackage{graphicx}
\usepackage{color}
\usepackage{amssymb}

\def\beq{\begin{equation}}
\def\eeq{\end{equation}}
\def\bea{\begin{eqnarray}}
\def\eea{\end{eqnarray}}

\newcommand*{\eqref}[1]{Eq.~(\ref{eq:#1})}
\newcommand*{\eqlab}[1]{\label{eq:#1}}
\newcommand*{\figref}[1]{Fig.~\ref{fig:#1}}
\newcommand*{\figlab}[1]{\label{fig:#1}}

\newcommand*{\seclab}[1]{\label{sec:#1}}

\def\VYP#1#2#3{{\bf #1}, #3 (#2)}  

\def\PLB#1#2#3{Phys.~Lett.~B~\VYP{#1}{#2}{#3}}

\def\PRD#1#2#3{Phys.~Rev.~D~\VYP{#1}{#2}{#3}}
\def\PRL#1#2#3{Phys.~Rev.~Lett.~\VYP{#1}{#2}{#3}}

\newcommand{\Omit}[1]{}

\begin{document}

\title{Detecting UHE Cosmics \& Neutrinos off the Moon; an
Optimal Radio Window. }

\author{O Scholten$^1$, J Bacelar$^1$, R Braun$^2$, A G de Bruyn$^{2,3}$,
H Falcke$^{2,4}$, B Stappers$^{2,5}$, and R G Strom$^{2,5}$}
\address{$^1$ Kernfysisch Versneller Instituut, University of Groningen,
9747 AA, Groningen, The Netherlands}
\address{$^2$ ASTRON, 7990 AA Dwingeloo, The Netherlands}
\address{$^3$  Kapteyn Institute, University of Groningen, 9747 AA, Groningen, The
Netherlands}
\address{$^4$ Department of Astrophysics,
IMAPP, Radboud University, 6500 GL Nijmegen, The Netherlands}
\address{$^5$ Astronomical Institute `A. Pannekoek', University of Amsterdam, 1098 SJ,
The Netherlands}
\ead{scholten@kvi.nl}

\begin{abstract}
We show that at wavelengths comparable to the length of the shower produced by
an Ultra-High Energy cosmic ray or neutrino, radio signals are an extremely
efficient way to detect these particles. Through an example it is shown that
this new approach offers, for the first time, the realistic possibility of
measuring UHE neutrino fluxes below the Waxman-Bahcall limit. It is shown that
in only one month of observing with the upcoming LOFAR radio telescope,
cosmic-ray events can be measured beyond the GZK-limit, at a sensitivity level
of two orders of magnitude below the extrapolated values.
\end{abstract}

\section{Introduction}

As an efficient method to determine the fluxes of UHE particles we are
investigating the production of radio waves when an UHE particle hits the moon.
Askaryan predicted as early as 1962~\cite{Ask62} that particle showers in dense
media produce coherent pulses of microwave \v{C}erenkov radiation. Recently
this prediction was confirmed in experiments at accelerators~\cite{Sal01} and
extensive calculations have been performed on the development of showers in
dense media to yield quantitative predictions for this effect~\cite{Zas92}. The
Askaryan mechanism lies at the basis of several experiments to detect (UHE)
neutrinos using the \v{C}erenkov radiation emitted in ice
caps~\cite{Mio04,Leh04}, salt layers~\cite{SALSA}, and the lunar regolith. The
pulses from the latter process are detectable at Earth with radio telescopes,
an idea first proposed by Dagkesamanskii and Zheleznyk~\cite{Dag89} and later
by others~\cite{Alv01}. Several experiments have since been
performed~\cite{Han96,Gor04} to find evidence for UHE neutrinos. All of these
experiments have looked for this coherent radiation near the frequency where
the intensity of the emitted radio waves is expected to reach its maximum.
Since the typical lateral size of a shower is of the order of 10~cm the peak
frequency is of the order of 3~GHz.

We propose a different strategy~\cite{Sch06} to look for the radio waves at
considerably lower frequencies where the wavelength of the radiation is
comparable in magnitude to the typical longitudinal size of showers. We show
that the lower intensity of the emitted radiation, which implies a loss in
detection efficiency, is compensated by the increase in detection efficiency
due to the near isotropic emission of coherent radiation. The net effect is an
increased sensitivity by several orders of magnitude for the detection of UHE
cosmic rays and neutrinos at frequencies which are one or two orders of
magnitude below that where the intensity reaches its maximum. At lower
frequencies the lunar regolith becomes increasingly transparent for radio
waves. This implies for the detection of UHE neutrinos that there are two gain
factors when going to lower energies; i) Increased transparency of the lunar
regolith already stressed in Ref.~\cite{Fal03}, and ii) Increased angular
acceptance, stressed in this work.

In Section {\bf 4} we discuss two specific observations, one for an existing
facility, the Westerbork Synthesis Radio-Telescope array (WSRT) where we have
observation time granted, and one for a facility which will be available in the
near future, the Low-Frequency Array (LOFAR).

\section{Model for Radio Emission}

There exist two rather different mechanisms for radio emission from showers
triggered by UHE cosmic rays or neutrinos. One is the emission of radio waves
from a shower in the terrestial atmosphere. Here the primary mechanism is the
synchrotron acceleration of the electrons and positrons in the shower due to
the geomagnetic field, called geosynchrotron
radiation~\cite{Fal03,Sup03,Hue05,Fal05,Ard05}. The second mechanism applies to
showers in dense media where the front end of the shower has a surplus of
electrons. Since this cloud of negative charge is moving with a velocity which
exceeds the velocity of light in the medium, \v{C}erenkov radiation is emitted.
For a wavelength of the same order of magnitude as the typical size of this
cloud, which is in the radio-frequency range, coherence builds up and the
intensity of the emitted radiation reaches a maximum. This process, known as
the Askaryan effect~\cite{Ask62} is the subject of this work.

The intensity of radio emission (expressed in units of Jansky's where
1~Jy~=~$10^{-26}$~W~m$^{-2}$Hz$^{-1}$) from a hadronic shower, with energy
$E_s$, in the lunar regolith, in a bandwidth $\Delta\nu$ at a frequency $\nu$
and an angle $\theta$, can be parameterized as~\cite{Sch06}
\bea
&&\hspace*{-1.5em} F(\theta,\nu,E_s)
 = 3.86 \times 10^4\; e^{-Z^2} \Big( {\sin{\theta}\over \sin{\theta_c}} \Big)^2
 \Big( {E_s \over 10^{20} \mbox{ eV} } \Big) ^2
 \nonumber \\ &&\hspace*{-1.5em}  \times
 \Big( {d_{moon} \over d } \Big)^2
 \Big( {\nu \over \nu_0 (1+(\nu/\nu_0)^{1.44})} \Big)^2
 ({\Delta\nu \over 100\mbox{ MHz}}) \; \mbox{Jy} \;,
 \eqlab{shower-l}
\eea
with
\beq
Z
 = (\cos{\theta} -1/n)
 \Big({n\over \sqrt{n^2-1}}\Big)\Big({180\over \pi \Delta_c}\Big)\;,
 \eqlab{Z}
\eeq
where $\nu_0=2.5$~GHz~\cite{Gor04}, $d$ is the distance to the observer, and
$d_{moon}=3.844 \times 10^8$~m is the average Earth-Moon distance. The angle at
which the intensity of the radiation reaches a maximum, the \v{C}erenkov angle,
is related to the index of refraction ($n$) of the medium,
$\cos{\theta_c}=1/n$. Crucial for our present discussion is the spreading of
the radiated intensity around the
\v{C}erenkov angle, given by $\Delta_c$ (in degrees). The $\sin{\theta}$ factor in
\eqref{shower-l} reflects the projection of the velocity of the charges in
the shower on the polarization direction of the emitted \v{C}erenkov radiation.
The dependence of $Z$, defined in \eqref{Z}, is suggested by working out some
specific cases, as presented in a following paragraph. For small values of
$\Delta_c$ it coincides with the formula found in much of the
literature~\cite{Zas92,Alv01,Gor04} however,
\eqref{shower-l} is more accurate for large spreading angles.

The spreading of the radiated intensity around the \v{C}erenkov angle,
$\Delta_c$, is, on the basis of general physical arguments, inversely
proportional to the shower length and the frequency of the emitted radiation.
Based on the results given in Ref.~\cite{Alv01} it can be parameterized as
\beq
\Delta_c = 4.32^\circ
 \Big({1\over \nu\,[\mbox{GHz}] }\Big)
 \Big({L(10^{20}\mbox{eV})\over L(E_s) }\Big) \;,
\eqlab{del_c_had'}
\eeq
where $L(E_s)$ is the shower length which depends on the energy. In
Ref.~\cite{Alv98}, values are given for the shower length (in units of
radiation lengths, equal to 22.1~g/cm$^{2}$ for lunar regolith). At an energy
of $10^{20}$~eV this corresponds to a shower length of approximately 1.7~m.

Since the angular spread of the intensity of the \v{C}erenkov radiation around
the \v{C}erenkov angle for the case $\lambda\approx L$ is crucial for our
considerations, we have derived an analytic formula for the angular
distributions for two different shower profiles following the approach given
in~\cite{Leh04}. For the first one, called the ``block'' profile, the number of
charged particles is constant over the shower length $L=L_b$, $\rho_b(x)=1$ for
$0<x<L_b$. This profile is not realistic for a shower as the full intensity
suddenly appears and disappears. For this reason we have also investigated a
second profile where the charge in the shower appears and disappears following
a sine profile, $\rho_s(x)=\sin{\pi x/L_s}$ with $0<x<L_s$.

\begin{figure}
    \includegraphics[height=7.9cm,bb=50 147 515 650,clip]{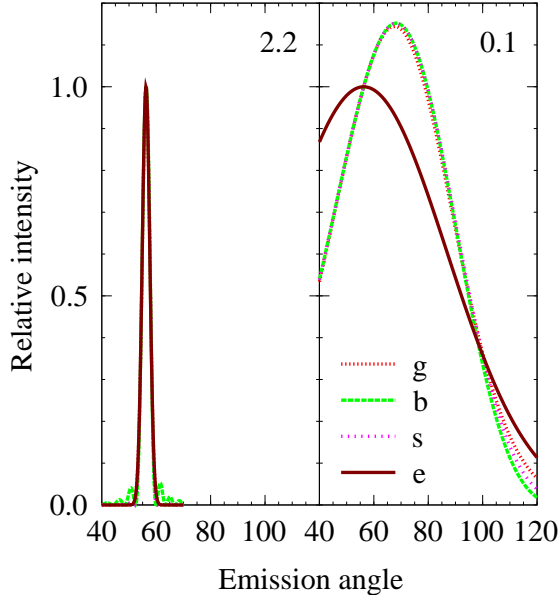}
\hspace{2pc} \begin{minipage}[b]{18pc}
\caption[fig9]{The angular spread around the \v{C}erenkov angle for different
shower-profile functions (see text) are compared to the parametrization used in
this work. The left (right) hand displays the results for 2.2 GHz (100MHz)
respectively. }
  \figlab{ang-spread}
\end{minipage}
\end{figure}

For the block longitudinal profile we reproduce the well known intensity
distribution found by Tamm~\cite{Tam39} for a finite length shower, normalized
to unity at the \v{C}erenkov angle, $ I_b(\theta) = \Big[ {\sin{\theta}\over
\sin{\theta_c}} {\sin{\pi \chi} \over \pi \chi} \Big]^2 $ with
$ \chi=(\cos{\theta} -1/n)L/\lambda $. For the normalized ``sine''
profile we obtain
\beq
I_s(\theta) = \Big[ {\sin{\theta}\over \sin{\theta_c}}
 {\cos{\pi \chi}\over (1-2\chi)(1+2\chi)} \Big]^2
\eqlab{I-sin}
\eeq
The predictions of these two formulas are compared (curves labelled 'b' for
block and 's' for sine in \figref{ang-spread}) with the exponential form used
in Monte Carlo simulations~\cite{Zas92,Alv01,Gor04}, $ I_e(\theta) =
e^{-((\theta-\theta_c)/\Delta_c )^2} $ with $\Delta_c$ is given by
\eqref{del_c_had'}. The comparison is made for a shower of $10^{20}$~eV.
To reproduce the angular spread of this calculation at 2.2~GHz we choose
$L_b=2.5$~m and $L_s=L_b \times 4/3=3.4$~m. The results are also compared to
those for a gaussian profile, derived in~\cite{Leh04}, $ I_g(\theta) =
\Big[ {\sin{\theta}\over \sin{\theta_c}} \Big]^2 e^{-Z^2} \;, $ with $Z$
given by \eqref{Z} (curve labelled 'g' in \figref{ang-spread}). From
\figref{ang-spread} it is seen that at 2.2 GHz the simple exponential form is reproduced
well by all three analytic forms. The block profile shows the well known
secondary interference maxima, due to the sharp edges of the profile, which are
not realistic for our case. Keeping parameters fixed the angular distributions
are now compared at 100 MHz (right hand panel of
\figref{ang-spread}). All three analytic forms agree quite accurately but differ
considerably from the exponential form. The reason for this difference lies
mainly in the pre-factor $\sin^2{\theta}$ which accounts for the radiation
being polarized parallel to the shower which does not allow for emission at
0$^\circ$ and 180$^\circ$. On the basis of the arguments given above we have
used the gaussian parametrization in \eqref{shower-l}.

The analytic formulas also work remarkably well at the quantitative level. At
an energy of $10^{20}$~eV the length of the shower is 1.7 m, as found in
Ref.~\cite{Alv98}. The value for the length used in
\eqref{I-sin} is $L_s=3.4$ m. For a sine profile the density of charged
particles exceeds 70\% of the maximum value (the definition of shower length)
for only half this distance (i.e.\ 1.7 m) which is in excellent agreement with
the shower length found the Monte-Carlo simulations.

Cosmic-ray-induced showers occur effectively at the lunar surface. For
neutrino-induced showers an energy-dependent mean free path has been used,
$\lambda_\nu= 130\; \Big( {10^{20} \mbox{ eV} \over E_\nu } \Big)^{1/3}$~km. As
a mean value for the attenuation length for the radiated power in the regolith
we have taken $\lambda_r= (9/\nu$[GHz])~m.

A crucial point in the simulation is the refraction of radio waves at the lunar
surface as was already stressed in Ref.~\cite{Zas92}. Due to internal
reflection at the surface the emitted radiation at high frequencies where the
\v{C}erenkov cone is rather narrow will be severely diminished. The major
advantage of going to lower frequencies is that the spreading increases,
allowing for the radiation to escape from the lunar surface. With decreasing
frequency the peak intensity of the emitted radiation decreases, however, the
peak intensity increases with increased particle energy. The net effect is that
at sufficiently high shower energies the aforementioned effect of increased
spreading is far more important, resulting in a strong increase in the
detection probability.

An additional advantage of using lower frequencies is that the sensitivity of
the model simulations to large- or small-scale surface roughness is diminished.
Since at lower frequencies already a sizable fraction of the radiation
penetrates the surface, its roughness will not make a major difference. This is
in contrast to high frequencies where most of the radiation is internally
reflected when surface roughness is ignored.

\section{Detection Limits}

\begin{figure}
\centerline{
    \includegraphics[height=7.9cm,bb=36 137 515 672,clip]{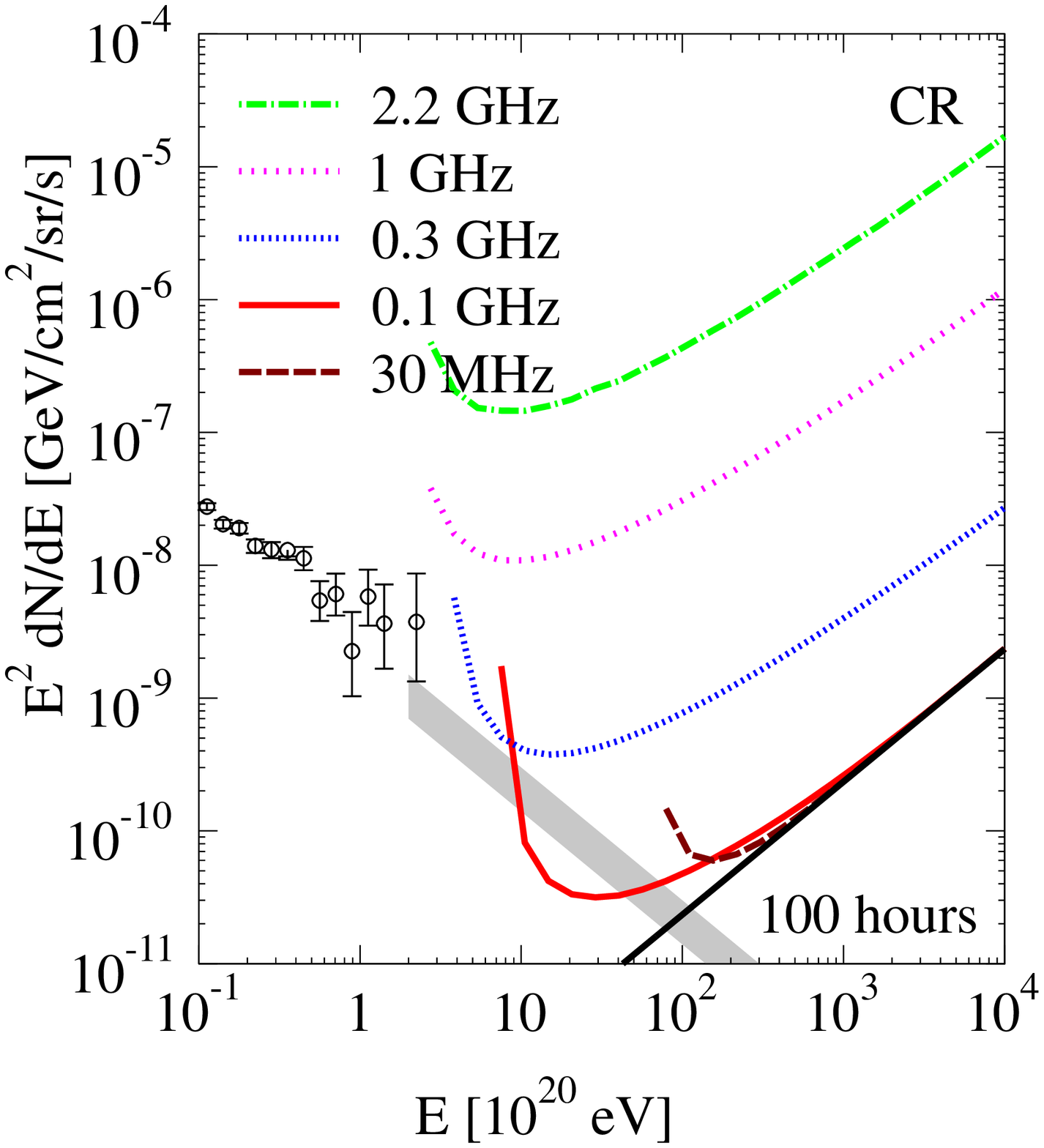}
\ \
    \includegraphics[height=7.9cm,bb=36 137 515 672,clip]{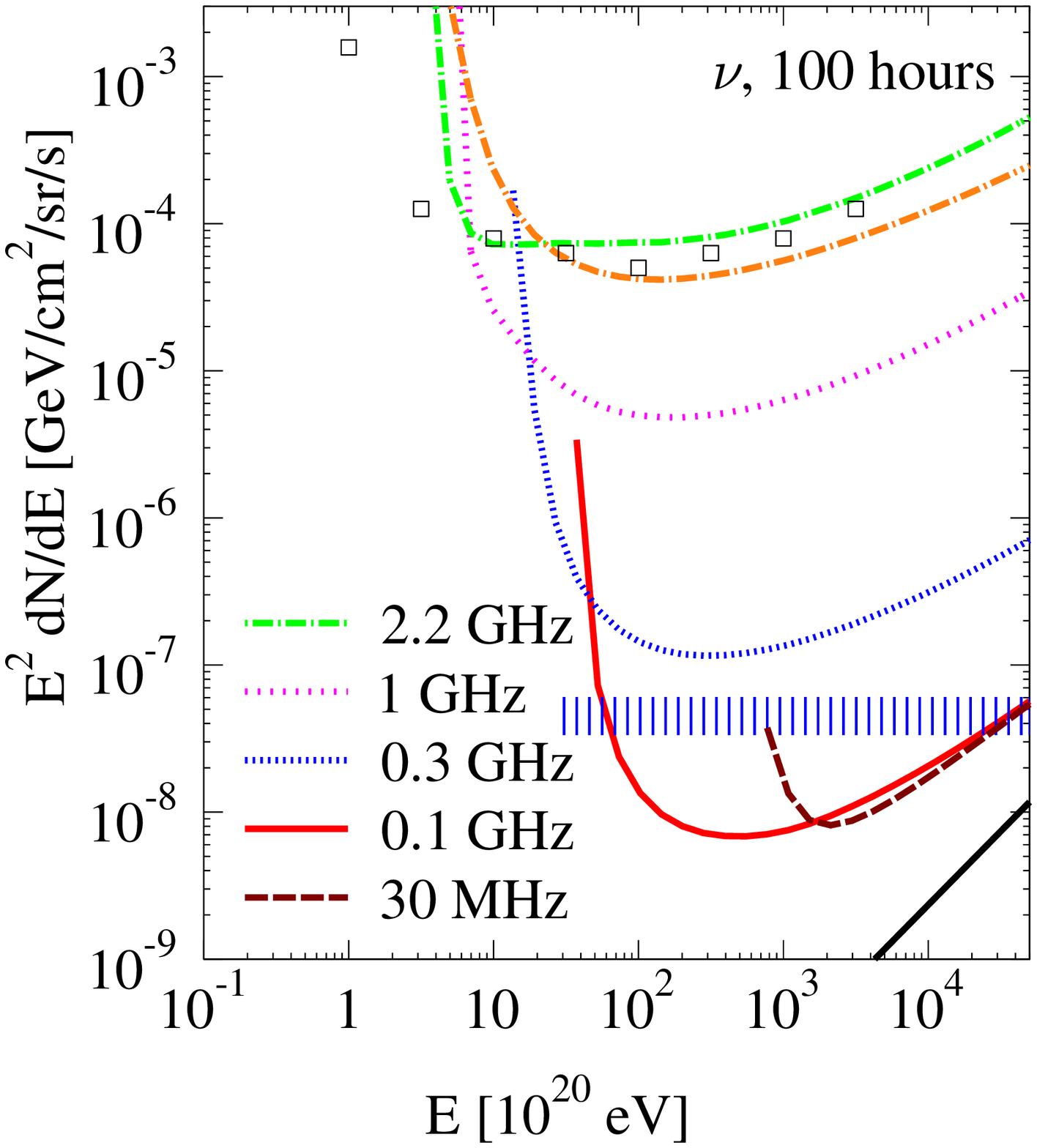}
}
\caption[fig3]{Flux limits (assuming a null observation) for cosmic rays (l.h.s.)
and neutrinos (r.h.s.) as can be determined in a 100 hour observation (see
text). In the curves for $\nu=30$~MHz a ten fold higher detection threshold is
used, corresponding to the higher sky temperature at this frequency. The points
given in the r.h.s.\ correspond to the data of the AGASA
experiment~\cite{Tak03}, the grey band is an extrapolation of the HiRes
data~\cite{Abb04}. The thick black line corresponds to the best possible limit
(vanishing detection threshold). The open squares for the neutrino flux are the
limits determined from the GLUE experiment~\cite{Gor04}.}
  \figlab{Flux}
\end{figure}

In \figref{Flux} the detection limits for UHE cosmic rays for different
radio-frequency ranges are compared with data from the AGASA~\cite{Tak03}
(points) and a linear extrapolation based on the data from the HiRes
experiment~\cite{Abb04} (grey bar). We regard an event to be detectable when
the power of the signal is 25 times larger than the ``noise level'' which we
take equal to $F_{noise}=20$~Jy, using a bandwidth of $\Delta\nu=20$~MHz,
values typical for LOFAR. It should be noted that for $\nu=30$~MHz there is a
strong increase in the sky temperature and we have used a ten-fold higher
threshold. As a result of the higher detection threshold the flux limit lies
considerably higher.

From the l.h.s.\ of \figref{Flux} one clearly sees that with decreasing
frequency one loses sensitivity for lower-energy particles. This follows
directly from
\eqref{shower-l} since with decreasing frequency the maximum signal strength
decreases and thus one exceeds the detection threshold only for more energetic
particles. If the energy of the cosmic ray is more than a factor 4 above this
threshold value the analysis presented in the previous section applies and the
detection limit improves rapidly with decreasing radio-frequency until one
reaches a frequency of 100~MHz where one obtains the optimum sensitivity.
Decreasing the frequency even lower provides no gain since the detection limit
has already reached the optimum given by the heavy black line in
\figref{Flux}.

In the r.h.s.\ of \figref{Flux} we compare the detection limits for UHE
neutrinos at different frequencies with the results obtained from the GLUE
experiment~\cite{Gor04}. For neutrino-induced showers only 20\% of the initial
energy is converted to a hadronic shower, while the remaining 80\% is carried
off by the lepton. This energetic lepton will not induce a detectable radio
shower. One sees similar trends as in the predictions for cosmic rays, in
particular the large gain in the determined flux limits with decreasing
frequency. At higher energies the limits for neutrinos do not increase as
steeply as those for cosmic rays. This is because the neutrino mean-free-path
decreases with energy, therefore increasing the probability for the neutrino to
initiate a shower close to the surface where the attenuation of the radio waves
is small. Our result at 2.2~GHz lies close to that of the GLUE experiment.

\section{WSRT \& LOFAR predictions\seclab{RealObs}}

The Westerbork Synthesis Radio Telescope (WSRT)~\cite{WSRT-l} consists of
fourteen 25~m parabolic dishes located on an east-west baseline extending over
2.7~km. Elements of the array can be coherently added to provide a response
equivalent to that of a single 94~m dish. Observing can be done in frequency
bands which range from about 115 to 8600~MHz, with bandwidths of up to 160~MHz.
The low frequency band which concerns us here covers 115-170~MHz. In tied-array
mode the system noise at low frequencies is $F_{noise}=$600~Jy. To observe
radio bursts of short duration, the new pulsar backend (PuMa II) will be used.
In the configuration which we propose to use, four frequency bands will observe
the same part of the moon with the remaining four a different section. In
total, coverage of about 50\% of the lunar disk can be achieved. We have
recently obtained 500 hours observing time of which 200 hours are granted for
the first year.

An even more powerful telescope will be the LOFAR array~\cite{LOFAR}. With a
collecting area of about 0.05~km$^2$ in the core (which can cover the full moon
with an array of beams), LOFAR will have a sensitivity about 25 times better
than that of the WSRT.  LOFAR will operate in the frequency bands from 30-80
and 115-240~MHz.

\begin{figure} 
\centerline{
    \includegraphics[height=7.9cm,bb=27 137 515 672,clip]{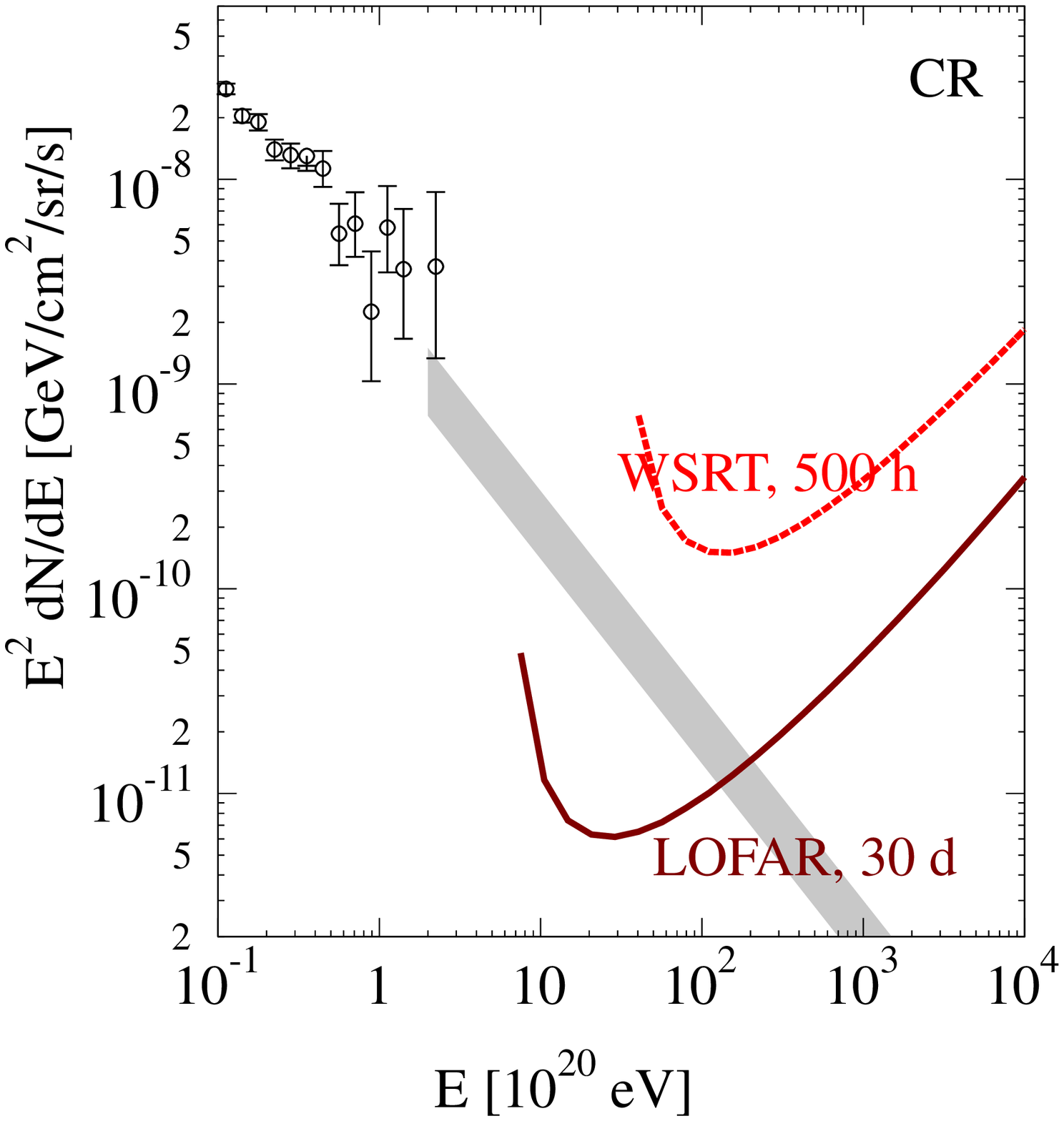}
\ \
    \includegraphics[height=7.9cm,bb=27 137 515 672,clip]{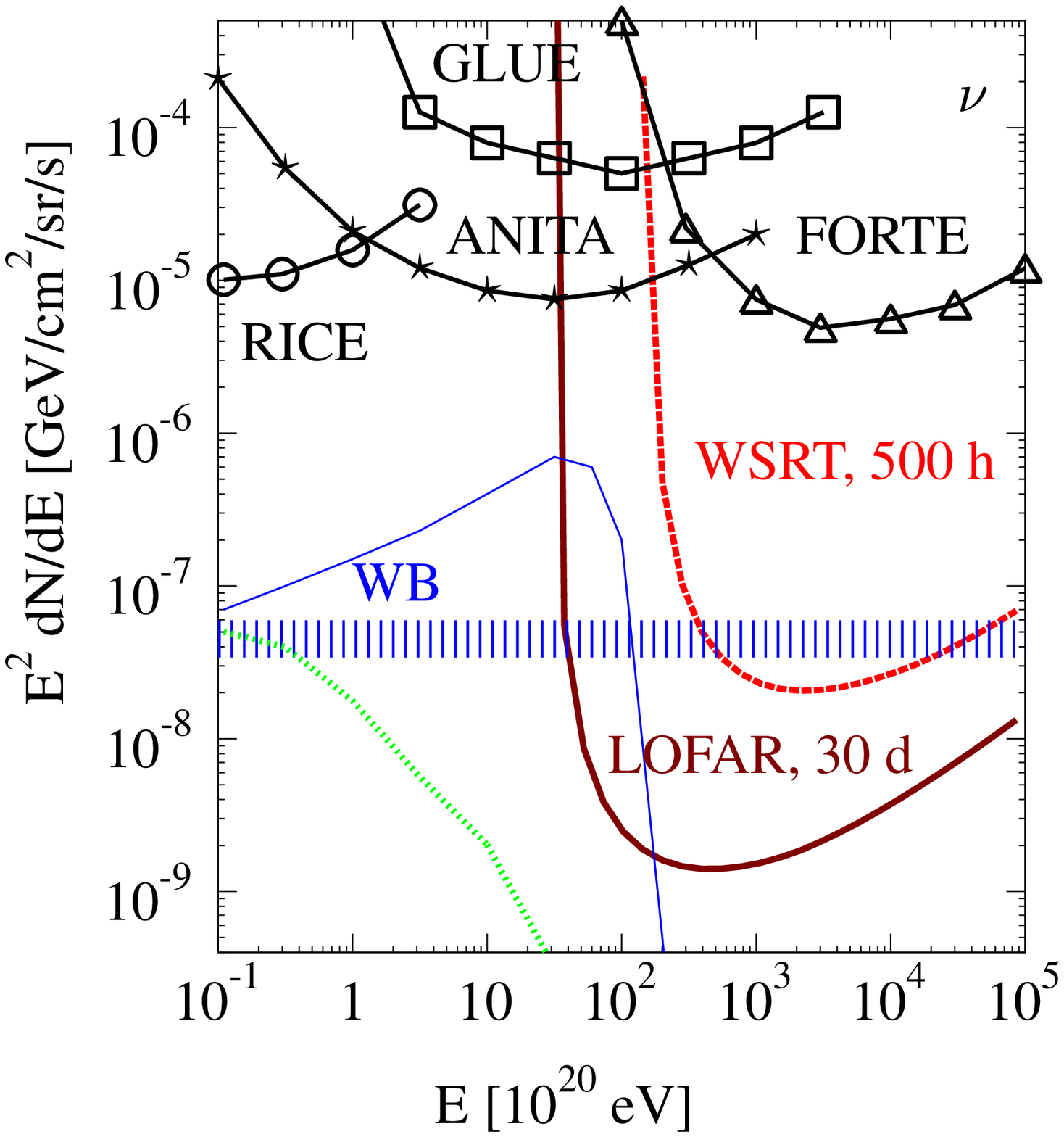}
}
\caption[fig5]{left: Flux limits on UHE cosmic rays (r.h.s.) and neutrinos (r.h.s.)
as can be determined in a 30 day observation with the LOFAR antenna system and
a 500 hour observation with WSRT. The data cosmic rays are the same as in
\figref{Flux}. The flux limits on UHE neutrinos are compared with various models,
in particular, WB~\cite{Bah01} (vertical bars), GZK~\cite{Eng01} (dotted thin
line), and TD~\cite{Sta04,Pro96} (solid thin line). The experimental limits on
the neutrino flux are from the RICE~\cite{Kra03}, GLUE~\cite{Gor04},
ANITA~\cite{Bar06}, and FORTE~\cite{Leh04}.}
  \figlab{LOFAR}
\end{figure}

Simulations show that a pulse of intensity 25$\times F_{noise}$, interfering
with the noise background, can be detected with $3
\sigma$ significance at a probability greater than 80\%. For this reason we
have assumed in the calculations a detection threshold of 25$\times F_{noise}$
for both the WRST and the LOFAR telescopes. A simulation for LOFAR, taking
$\nu=120$~MHz, bandwidth of $\Delta\nu=20$~MHz, a signal-detection threshold of
500~Jy,  and an observation time of 30 days is shown in \figref{LOFAR} for
cosmic rays and neutrinos. The results are compared with the limits that can be
obtained from a presently proposed observation for 500 hours at the WSRT
observatory assuming a detection threshold of 15,000~Jy, $\nu=140$~MHz,
bandwidth of $\Delta\nu=20$~MHz, and a 50\% Moon coverage.

In \figref{LOFAR} the flux limits are compared to the predictions of several
models and with those of other experiments. With the existing WSRT a limit on
the neutrino flux can be set which falls just below the WB bound. However, even
this will constrain different top-down scenarios, discussed in the literature.
With the proposed LOFAR facility this limit can be improved considerably to
reach, for the first time, a limit well below the WB bound for neutrinos. In
addition one has a good chance to see evidence (in only a 30 day period) of
cosmic ray events at an energy one order of magnitude higher than presently
observed.

\section{Summary}

We have demonstrated~\cite{Sch06} the clear advantage of using radio waves at
frequencies well below the \v{C}erenkov maximum. The optimum frequency will be
that where the length of the shower, of the order of several meters in the
lunar regolith, is of the same order of magnitude as the wavelength of the
radio waves where the radio-emission pattern is more isotropic. The gain in
efficiency at lower frequencies is such that with the upcoming LOFAR facility
one can seriously investigate realistic top-down scenarios for UHE neutrinos
and be sensitive to neutrino fluxes well below the Waxman-Bahcall limit. Even
now with the existing WSRT, profiting from its capability to measure right in
the radio-frequency window where the detection efficiency is highest, one is
able to set limits on neutrino fluxes orders of magnitude below the present
limit in only a 500~h observation period.

For UHE cosmic rays the LOFAR facility offers, because of the availability of
an optimal radio-frequency window, a very powerful tool to determine the flux
beyond the GZK limit. In only a 30~day observing period one is sensitive to a
flux which is more than one order of magnitude below the extrapolation of the
measured flux from below the GZK limit.

\ack
This work was performed as part of the research programs of the Stichting voor
Fundamenteel Onderzoek der Materie (FOM) and of ASTRON, both with financial
support from the Nederlandse Organisatie voor Wetenschappelijk Onderzoek (NWO).

\end{document}